# DesignCon 2011

# High Speed Parallel Signal Crosstalk Cancellation Concept

Chad M. Smutzer, Mayo Clinic
smutzer.chad@mayo.edu

Michael J. Degerstrom, Mayo Clinic
degerstrom.michael@mayo.edu

Barry K. Gilbert, Mayo Clinic
gilbert.barry@mayo.edu, 507-284-4056

Erik S. Daniel, Mayo Clinic

## Abstract

High performance computing (HPC) systems make extensive use of high speed electrical interconnects, in routing signals among processing elements, or between processing elements and memory. Increasing bandwidth demands result in high density, parallel I/O exposed to crosstalk due to tightly coupled transmission lines. The crosstalk cancellation signaling concept discussed in this paper utilizes the known, predictable theory of coupled transmission lines to cancel crosstalk from neighboring traces with carefully chosen resistive cross-terminations between them. Through simulation and analysis of practical bus architectures, we explore the merits of crosstalk cancellation which could be used in dense interconnect HPC (or other) applications.

## Author's Biographies

**Chad M. Smutzer** received a BSEE from the University of Iowa in Iowa City. He is currently a Senior Engineer at the Mayo Clinic Special Purpose Processor Development Group where he performs signal and power integrity analysis.

**Michael J. Degerstrom** received a BSEE from the University of Minnesota. Mike is currently a Senior Engineer at the Mayo Clinic Special Purpose Processor Development Group. His primary area of research and design has been in the specialty of signal and power integrity.

**Barry K. Gilbert** received a BSEE from Purdue University (West Lafayette, IN) and a Ph.D. in physiology and biophysics from the University of Minnesota (Minneapolis, MN). He is currently Director of the Special Purpose Processor Development Group, directing research efforts in high performance electronics and related areas.

**Erik S. Daniel** received a BA in physics and mathematics from Rice University (Houston, TX) in 1992 and a Ph.D. degree in solid state physics from the California Institute of Technology (Pasadena, CA) in 1997. He is currently Deputy Director of the Special Purpose Processor Development Group, directing research efforts in high performance electronics and related areas.

# Introduction

It is well known that a single transmission line with a well-defined characteristic impedance may be terminated with a single resistor of the same value as the impedance, thereby eliminating reflections and providing an accurate replica of the input signal at the output. It is much less widely known (but represented in the literature [1]) that, in principle, it is possible to transmit essentially noise-free signals along a set of uniform coupled transmission lines that are terminated with a well-defined characteristic impedance matrix. As will be described in this paper, terminations consisting of a network of resistances between signal lines, and between these same signal lines and a reference voltage, can be provided to reduce or eliminate reflections, canceling the signal crosstalk (i.e. coupling of energy from one conductor to a nearby conductor) that would otherwise have been present. In principle, the use of a crosstalk cancellation network would allow nearby signaling lines to be packed together much more tightly, allowing a greater number and density of parallel point-to-point links to be used within a given package or printed circuit board footprint. To our knowledge, this concept has never been demonstrated in practice.

We begin this paper with a brief theoretical discussion of a specific crosstalk cancelling method. We then evaluate the impact of this method on packaging density and various other figures of merit in the context of a typical HPC design. Next we apply the crosstalk cancellation signaling concept in a series of notional HPC system time-domain simulations using realistic active and passive interconnect models. Within the simulation section, we explore a number of variations to the interconnect components and buffer stimulus while quantifying the impact to the vertical eye diagram opening. The paper is completed with conclusive remarks and suggestions for additional effort.

## **Crosstalk Cancellation Basic Principles of Operation**

For the purposes of this discussion, it is assumed that the reader has a fundamental understanding of crosstalk and its impact to signal integrity in a traditional HPC system, as well as basic transmission line theory. A non-exhaustive discussion of crosstalk and the essential modeling and simulation principles can be found in [2].

Crosstalk is typically managed by noise budgeting. However, there is a not well known theoretical approach to appropriately terminating tightly coupled transmission lines. A brief theoretical treatment is reproduced in Appendix A.

One can think of the mechanism of crosstalk cancellation as simply driving a coupled transmission line (wire bundle) having impedances that vary depending on the state of the signals through the individual wires. For this case the crosstalk is cancelled if the bundle is properly terminated by resistors of values computed in the Zt matrix (to be described below). For simplicity such a methodology is simply referred to as "crosstalk cancellation" throughout this paper. As long as a transmitter (output buffer) is capable of driving wires that have state-dependent characteristic impedances, then signals are transmitted and received undisturbed, assuming lossless media, from transmitter to receiver (input buffer).  In theory, terminations are not restricted to be placed only at the receiver.  However, we suggest it may not be practical to alternatively provide termination at the transmitter between the voltage source and the supply rail (as would be required) or to use similar techniques, such as those suggested by Broyde (See Fig. 2 in [3]).

To quickly illustrate the procedure studied in this work to employ crosstalk cancellation, and to illustrate some key features, the simple example of Figure 1 is employed.  In this example, a 2-wire bundle is utilized where a second wire is placed directly above a first wire in what is known as a broadside stripline configuration.  Broadside striplines are often used for differential signaling but in this case the two wires are driven single-ended.

The design flow is as follows.  First a wire bundle is proposed to meet performance criteria and also be easily manufactured.  Next, a 2-dimensional cross-section is analyzed by an electro-magnetic tool, such as Synopsys HSPICE 2D field solver, to produce both inductance (L) and capacitance (C) matrices that are used by Matlab code to compute resistive termination matrices.  The same cross-section is used to create a w-element transmission line model to provide an electrical representation of the wire bundle with a finite length prescribed by the designer.  The Synopsys HSPICE 2D field solver has the advantage that the L/C matrices are computed and then the w-element coupled transmission line model is automatically computed.  Next the termination matrix is applied to the receiver side and sources are applied to the driver side of the w-element.  The sources used in this example are ideal.  In a more realistic simulation (such as those presented later), transistor-based sources would replace ideal sources and packaging and other models would be added to the full link model.

In the lower right corner of Figure 1 the receiver voltages and currents are plotted for both wires.  The wire voltages have no crosstalk even though the wires are strongly coupled.  Crosstalk is cancelled since the sources supply either +/-6 or +/-12.5 mA depending on the states of the wires.  In this example, the wires are symmetric, i.e., both wires have the same self-inductance and capacitance, but this is not a requirement for crosstalk cancellation.  As the number of wires increases, the number of finite current states and number of resistors in a complete termination network increase as well.  Collectively, the wire-bundle and transmitting and receiving circuitry (including termination) is more accurately described as "crosstalk cancellation signaling."

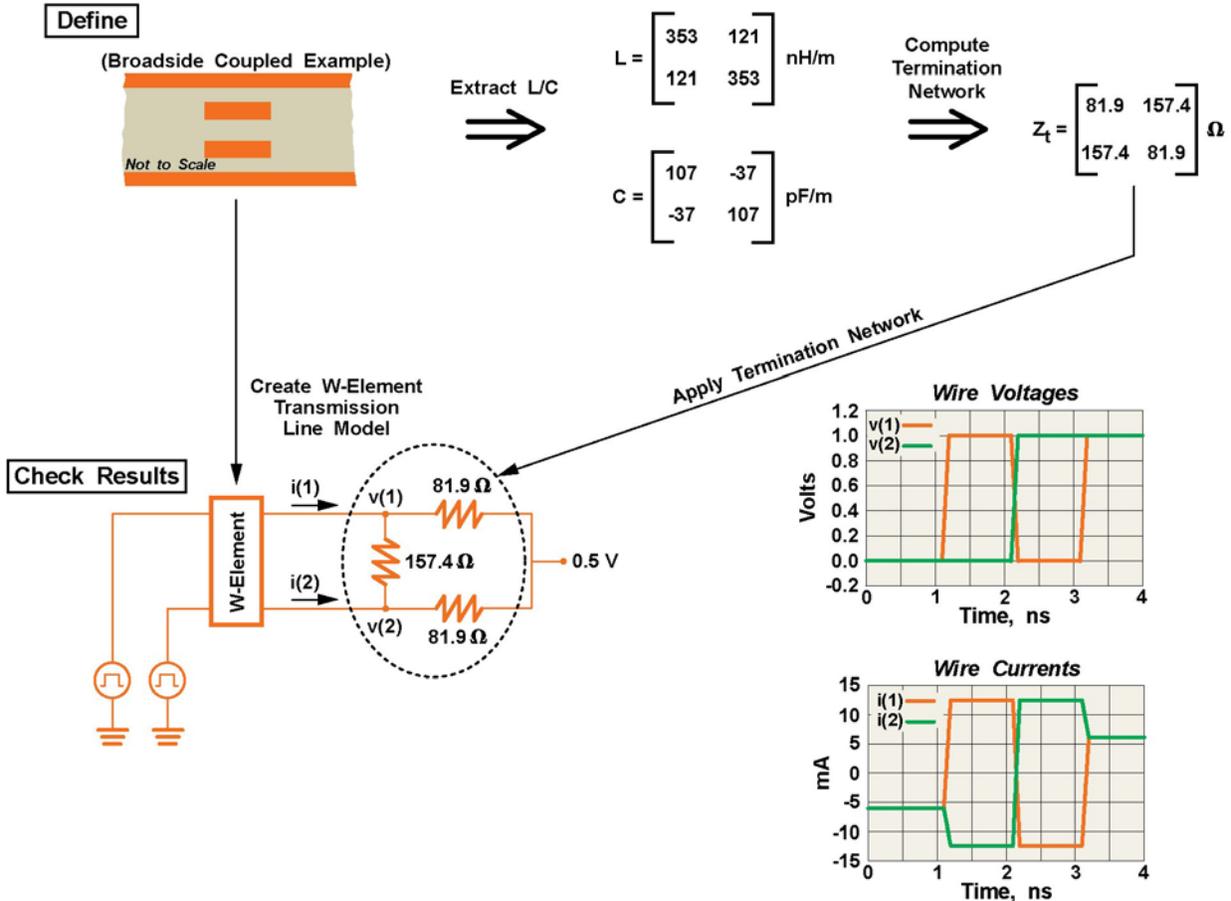

**Figure 1: Simple Crosstalk Cancellation Example (41490v2)**

For best results an ideal voltage source with no source resistance is desired to drive a wire bundle with tight coupling. Except for rare cases, circuitry that approximates a voltage source with low source resistance, such as the bipolar emitter follower circuit, is not available. Optionally, it may be possible to use a CMOS buffer with sufficient parallel PMOS pull-up and NMOS pull-down devices such that the source resistance is sufficiently low. Such an approach results in voltage swings nearly rail to rail and hence are not attractive when trying to meet I/O power budgets. Nevertheless, to illustrate nuances of crosstalk cancellation design and to compare results against conventional design, the highly parallel push-pull buffer was utilized for analysis. As will be discussed later, when the results of these simulations are described, we did find that finite source impedance drivers did present some limitations in crosstalk cancellation signaling performance. Therefore, we will continue in this section to briefly discuss alternate driver approaches, even though these were not further addressed through the course of this study.

Ideal crosstalk cancellation requires accurate termination between all wires and a termination voltage, as well as between pairs of wires. Note that we refer to these termination resistors later in this paper as self and cross termination resistors, respectively. However, it is not clear whether termination resistance accuracy is more important for crosstalk cancellation when compared with the accuracy required for a traditional signaling/termination approach. The value of the desired termination resistances for crosstalk cancellation is subject to variability – possibly more so than that of a traditional design (which will be discussed later in the context of the effects of PWB layer-to-layer shifting.). Therefore it is likely that crosstalk cancellation links would require some sort of power-up training sequence to set adjustable termination values.

Although not pursued through the course of the simulations undertaken in this study, we note that if the capability to automatically adjust terminations is provided, then it is likely that a similar technique can be implemented in which a low-resistance voltage source driver can be replaced by a code-dependent current source driver as described in [4]. Such a driver pulls or pushes current according to the logic states prescribed for each of the wires.

## Example Crosstalk Cancellation Network and Associated Figures of Merit

To gain insight into the characteristics of a tightly coupled wire bundle intended for use with crosstalk cancellation, a 12-wire bundle was analyzed. This 12-wire bundle, shown in the upper right corner of Figure 2 has ¼ oz. thick metal layers with .004" wide striplines and .004" edge-to-edge spacings. The dielectrics are .004" thick. The dielectric constant was arbitrarily chosen as 3.0.

The wire bundle geometry was defined in Synopsys' HSPICE 2D field solver to compute the resistance, inductance, conductance and capacitance (RLGC) matrices. The L and C matrices were analyzed to compute Zt, the termination matrix. For 12 wires, there are 12 self-terminations (between each wire and a reference voltage), and 66 cross terms (terminations between the wires) for a total of 78 terminations.

The termination values vary greatly in this example, ranging from 110Ω to more than 14 kΩ. The 110 Ω resistors map to the four conductors closest to the reference plane. The 224 Ω resistors map between reference planes and the outer conductors that are on the layers with three conductors. For a listing of resistor values for the 12-wire bundles see Table 1 in Appendix B. To plot a readily viewable histogram of the termination values, the conductance values are plotted in Figure 2 rather than the resistance values. Since resistor values are actually quite high compared to 50 Ω, circuit implementation is easier. For example, the switch enabling each resistor in a parallel network can be physically small since the higher switch resistance associated with a small switch adds little additional resistance to higher-valued resistors.

With many of the 78 termination values being very large-valued resistors, it seems feasible that many of the high-valued resistors need not be implemented. In fact, some wires may not even require termination to a supply, terminating only to nearby wires, if these nearby wires provide a more strongly coupled reference than that of the conventional reference planes. As will be discussed below, we did consider simulated cases in which the termination matrix was significantly reduced / simplified. The details of this matrix reduction will be discussed below, but here we will discuss one related theoretical consideration. Cross termination resistors supply current between wires that can switch out of phase from one another whereas self termination resistors switch current from wires into or out of a fixed reference voltage supply. Therefore the cross termination resistors can (in out-of-phase situations) carry twice the current than that of the self termination resistors for the same resistance values.  Hence, when considering elimination of resistances above some cutoff value, cutoff values differing by a factor of two should be used when considering self and cross termination resistors.  For example, if self termination resistors are implemented for values of 250 Ω and below, then cross termination resistors should be implemented for values of 500 Ω and below.

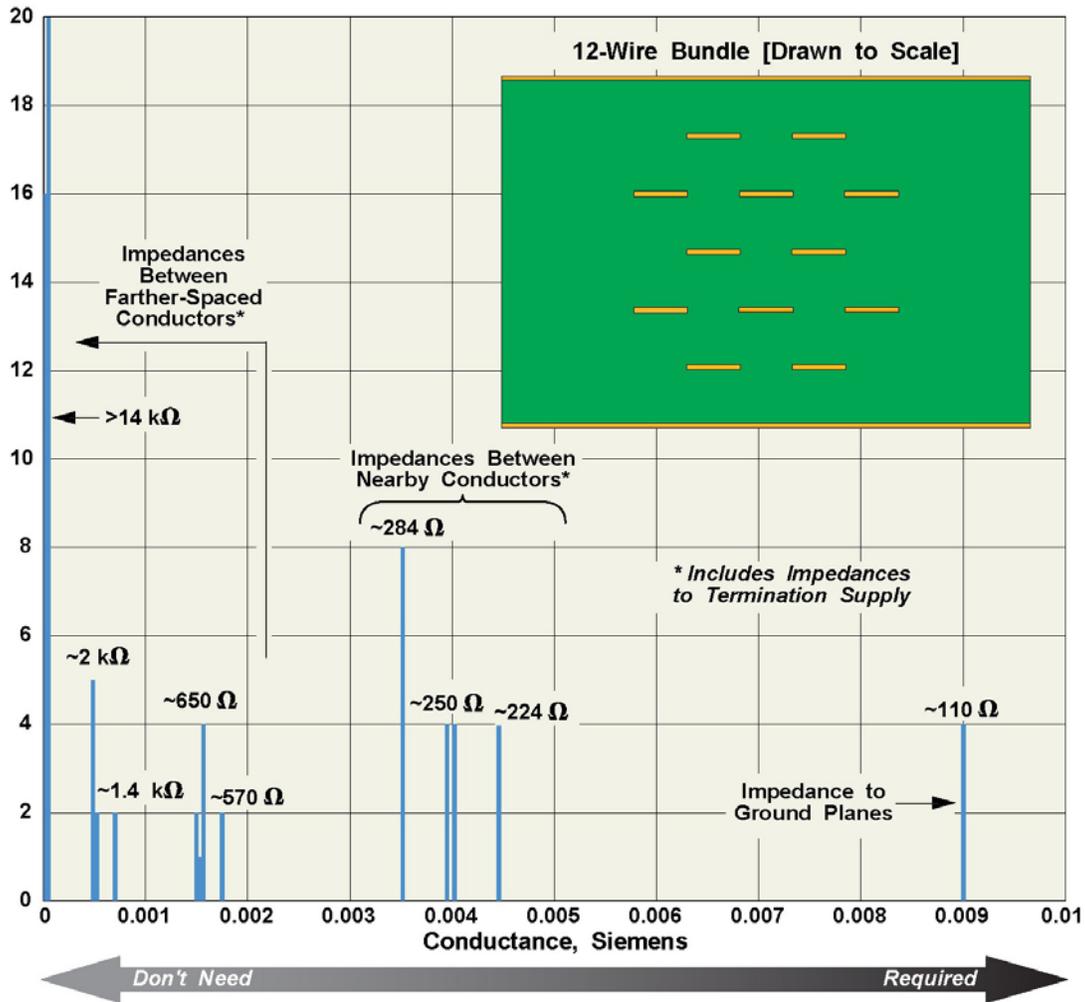

**Figure 2:  Histogram Showing Range of Impedances for Example 12-wire Bundle (41544v2)**

For simulation comparison purposes, it was necessary to create a model of a more traditional wire bundle with modest coupling which could be compared with the highly coupled 12-wire model.  In Figure 3 a more traditional 6-wire bundle is compared to the same 12-wire bundle discussed above.  Note that the 6-wire bundle represents a more traditional design where wires are adequately spaced and reference planes are utilized to keep crosstalk to acceptable (though in this case, still detectable) levels (since the crosstalk is not cancelled in a traditional signaling scheme).  Also note that the 12-wire bundle has the identical PWB layer build as that of the 6-wire bundle in terms of layer count and dielectric thicknesses, such that these two structures occupy the same cross-sectional area.  In fact the 12-wire bundle simply replaces the inner two reference planes of the 6-wire bundle with three striplines at each of these layers.  In the analysis sections of this paper, it is shown that these two bundle designs are selected as competing approaches that can be routed between 1 mm pin rows.

Although detailed simulations with each of these models are performed below, the table in the middle of Figure 3 compares simple performance characteristics between the two wire bundle designs. These comparison data are based on an assumed buffer with ideal voltage sources having no series resistance driving a lossless wire bundle. For the 6-wire bundle the wires are assumed to be terminated with 50 Ω resistors to 0.5 V. The 12-wire bundle is similarly terminated with resistors to 0.5 V and with resistors between wires with the resistor values set according to values computed in the Zt matrix.

The average bundle current is the average value of the current (over all the possible logical codes) flowing into the 0.5 V termination supply. Actually, the values that are reported are the average of the absolute value of this current since the termination supply will source current when enough wires are at logic low and will sink current when enough wires are at logic high. Since the 12-wire bundle, depending on the logic states, will return much of the current in individual wires through other signal wires (as opposed to through ground reference traces), then the average bundle current flowing into the termination voltage is only 8.1 mA. For the 6-wire bundle, essentially all the current flowing into the bundle flows through the 50 Ω termination resistors into the 0.5 V termination supply. It is easy to understand that the 12-wire bundle, with only four wires having 110 Ω and four additional wires of 224 Ω to the termination supply, would have significantly less average bundle current than that of the 6-wire bundle that has six wires with 50 Ω to the termination supply.

In addition to the average bundle current, the maximum bundle current is also computed. The maximum bundle current is a very important parameter since higher bundle currents create more simultaneous switching current which, along with inductive packaging parasitics, create more power supply noise. The 12-wire bundle has just over one-half the 6-wire maximum bundle current. The maximum bundle current for the 6-wire bundle occurs when all wires switch in phase and is a well understood problem by signal integrity engineers who manage simultaneous switching noise (SSN). The maximum bundle current for the 12-wire bundle is not as evident. However, if all wires switch in phase, then inner wires switch very little current. For those wires that do switch current, they do so into terminations significantly higher than 50 Ω. It is important to understand that these switching current values represent current that flows into a termination supply. In a push-pull driver implementation the current values represent the sum of the current from VDD and VSS. Therefore, effective current cancellation at the driver implies that VDD and VSS supplies are adequately coupled to one another by either using very low inductance packaging or by using an adequate amount of on-chip VDD to VSS decoupling capacitance.

Even though bundle currents for a 12-wire bundle are low, the maximum current in a single wire was computed to be 18.2 mA compared to 10.3 mA for that of the 6-wire bundle. If the wires in the 6-wire bundle were completely isolated from one another each wire would have +/- 10.0mA in each logic state (0.5 V across 50 Ω). Accordingly, the maximum 6-wire bundle current would be 60.0 mA. A wire in the 12-wire bundle can reach a peak value when an inner wire switches out of phase from all surrounding wires. For the case when a single wire has peak switching current then the bundle current will be far from its maximum value, since the high current in the wire with the peak current will cancel with currents in nearest-neighbor wires. Although peak currents within one wire may create localized crosstalk, peak bundle currents, which set the SSN levels, are expected to have a much larger impact on overall noise.

The final line in the table of Figure 3 shows the average power dissipated in the termination matrix. The termination matrix in the 12-wire bundle dissipates almost twice the power as that for the 6-wire bundle. For the 12-wire bundle much of the power is dissipated in resistors terminating between the wires. Assuming that both wire bundles can operate to equal maximum data rates, then the 12-wire bundle will have twice the I/O bandwidth compared to that of the 6-wire bundle. Therefore I/O used to drive the 12-wire bundle would offer slightly better energy efficiency in terms of mW/(Gb/sec) than that of traditional I/O used for the 6-wire bundle. Crosstalk cancellation energy efficiency would also decrease since additional power will be used by the required training and coding blocks described above. Conversely, it may be possible to remove the two sets of three-abreast wires in the 12-wire bundle to reduce the coupling strength and thereby reducing currents required between wires such that overall power efficiency can be improved while matching the I/O bandwidth of the traditional 6-wire bundle. Such an approach also lowers PWB metal layer counts. Many trade-offs such as these can be studied to tailor crosstalk cancellation technology to meet certain performance, cost, and other objectives.

The lower half of Figure 3 depicts the value of the current in each wire at all $2^n$ logic states, where n is the number of wires in the bundle. The 6-wire bundle has only 64 logic states and there are only four unique current values depending on the logic states. As mentioned above, if the wires had no coupling the only current states would be either +10 or -10 mA. The 12-wire bundle has 4096 logic states with considerably more variation of current across each wire depending on wire location and the logic states. While it is difficult to glean many of the figures of merit (as shown in the table) from these charts, it does offer a better visual interpretation of the behavior of current in traditional versus crosstalk cancellation approaches.

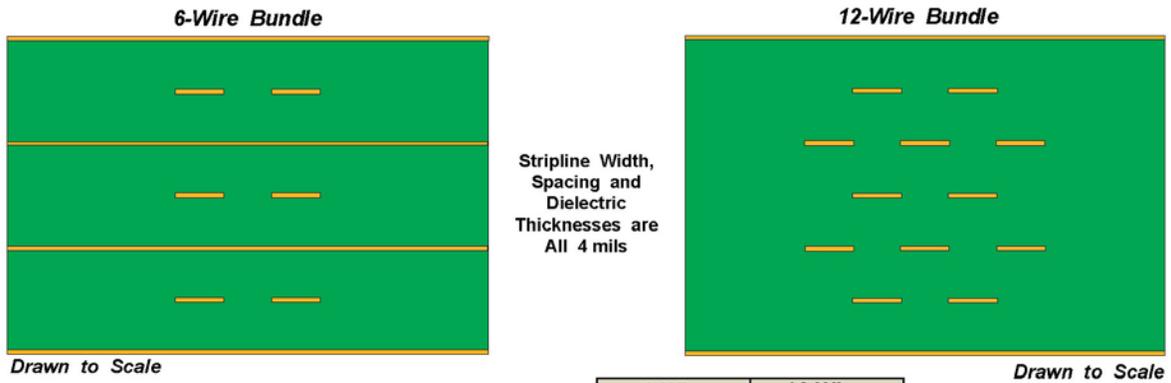

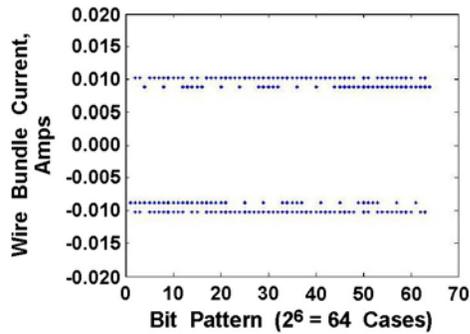
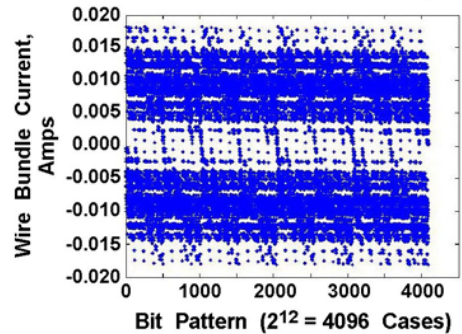

**Figure 3: Important Wire Bundle Characteristics (41495)**

These 12-wire and 6-wire bundle models will be carried forward into the more detailed simulations described in the following sections.

## Simulation Environment and Associated HPC System Assumptions

The previous sections of this document discussed how uniform coupled transmission lines (or wire bundles) could be terminated such that crosstalk was eliminated. Unfortunately, a typical end to end high-speed link is far from uniform and can include discontinuities and uncoupled regions in structures such as via pin fields or package breakout areas. It is in these non-uniform regions that resistive cancellation may not be as effective. In addition, as discussed above, finite output buffer impedance is expected to impede crosstalk cancellation signaling to some degree. In this section we explore the application of crosstalk cancellation in a more realistic HPC environment using time-domain simulation, incorporating such features as realistic output buffers, realizable pin field breakout patterns, and complete lossy transmission line models.

## Link Architecture

The general plan for evaluating crosstalk cancellation signaling was to create a simplified high-speed link for two variations of passive transmission media (discussed above); a traditional 6-wire design and a more densely populated 12-wire configuration to which the crosstalk cancellation signaling concept will be applied, along with a number of realistic, non-ideal components. The simulation link shown at the top of Figure 4 includes a series of high-speed transmit buffers, a non-uniform TX and RX pin field, a uniform transmission line structure and a resistive termination network. Each of these main blocks in the link architecture is discussed individually in following sections. Additionally, variations on each of these general structures are discussed in the results section below. This link model is simplified in that it does not include package or die parasitics, connectors, or any other passive components that may be present in a complete link.

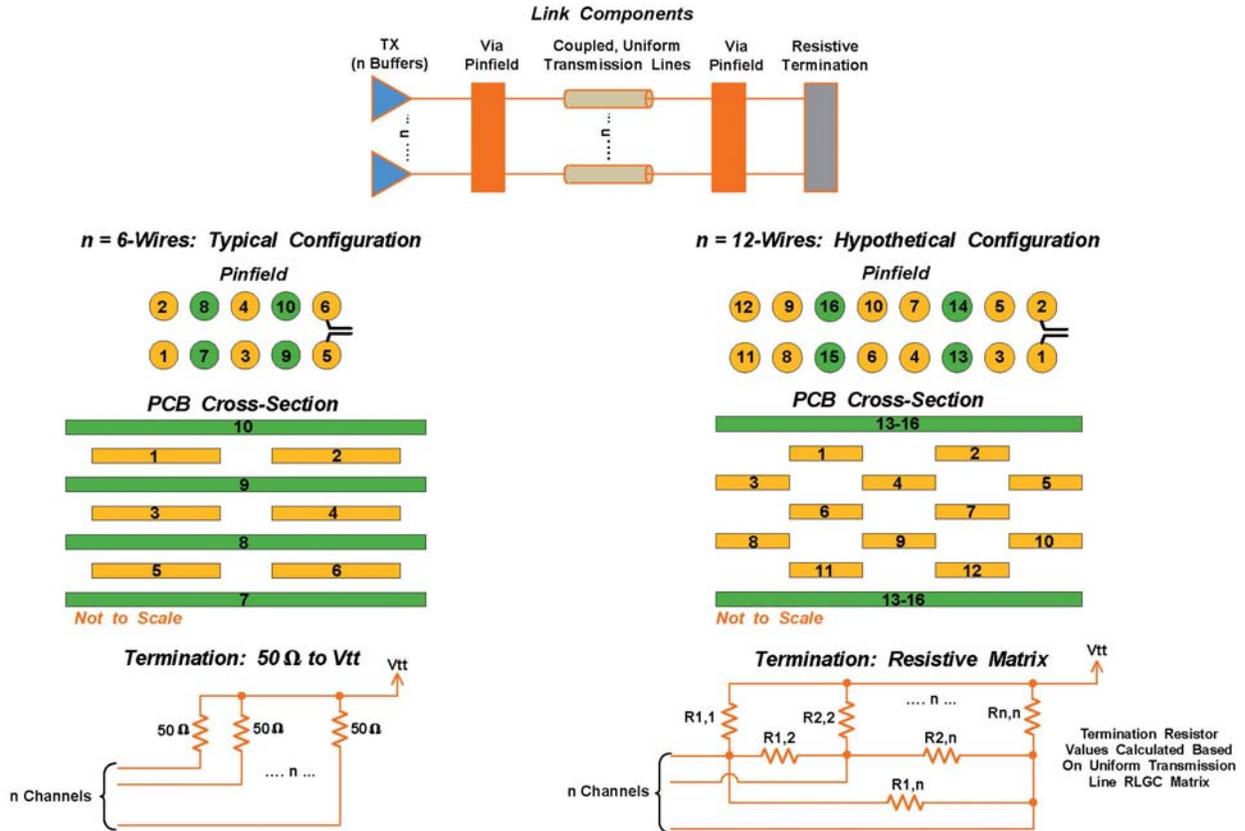

**Figure 4: Simulation Link Components and Configuration for 6-Wire and 12-Wire Bundles (41491v2)**

## Link Architecture: Transmit Buffer

The link begins with a transistor model for a push-pull style, source-series terminated (SST) output buffer. This transmitter nominally has a 50 Ohm series output impedance by design and operates from 0 to 1 VDC full swing. The choice of this buffer was influenced primarily by prior experience [5], and because of the need for high-speed capability to stress the frequency-dependent channel. The SST architecture was emulated using 90nm CMOS transistor models downloaded from MOSIS [6]. Transistor parameters were adjusted so as to nearly match the performance (voltage and timing) of the specific SST model as designed in [5].

While desirable for a traditional link design, the (relatively large) 50 Ω output impedance is actually not ideally suited for the crosstalk cancellation concept, as discussed above. Alternate buffer schemes were considered, however, the effort required to investigate various possible buffer schemes and implement them for simulation was out of scope. While impractical from a physical CMOS-based implementation perspective, a voltage source with a low (ideally zero) output impedance is a viable alternative to use for these simulations. Therefore, the SST buffer was modified by effectively placing many copies of the 50 Ω buffer in parallel to reduce the output impedance to nominally 1.67 Ω. This value was selected based on simulation results (discussed below, shown in Figure 11) showing that the eye diagram opening approaches a maximum with approximately 1.67 Ω source impedance. Towards the end of this paper, we will discuss the impact of finite buffer impedance on crosstalk cancellation signaling performance.

## Link Architecture: Transmission Lines

A not-to-scale drawing of the traditional 6-signal PWB cross-section (discussed above) is shown on the left side of Figure 4. By removing the reference planes between the signal layers, 12 signal wires with the same trace width and spacing can be routed using roughly the same 2D cross-sectional area as seen in the true-scaled depictions of the cross-sections in Figure 3.

The uniform transmission line structures were modeled using HSPICE's 2D field solver tool with the resulting RLGC model imported to Agilent's Advanced Design System (ADS) time-domain solver. The cross-section dimensions and material properties were discussed above. The resulting odd- and even-mode impedance between adjacent wires on the same layer in the 6-wire bundle was 47 Ω and 55 Ω, respectively. Note that in these simulations, the wires were not driven differentially – the wires were driven with single-ended SST buffers. The odd- and even-mode impedances are quoted here mainly as an indication of the (significant, but as we will see below, acceptable) coupling between the two wires in each layer of the 6-wire bundle. In our opinion, this level of coupling would be considered aggressive for a system design with no crosstalk mitigation, but is not unrealistic as a comparison to the 12-wire bundle as is discussed in the next section in the context of a 1 mm pin pitch breakout region. A similar discussion of even- and odd- mode impedance has little relevance for the 12-wire bundle where each wire has several neighbors with strong coupling – the strength of coupling in this case, however, is reflected in the magnitude of the terms in the resistive termination (e.g., as shown in Table 1). In each of the links, the length of transmission line used was 4".

## Link Architecture: Via Pin Field

The nominal pin field models, shown in Figure 5, were created in ANSYS HFSS using a tool developed by the Mayo SPPDG known as PinBuilder. The transmission line geometries defined in the 2D field solver were repeated for the striplines feeding the pins in this 3D model. For the majority of the simulations, the row and column pitch in the via pin field was 0.5 mm (20 mil) and 1 mm (40 mil), respectively, as depicted in the figure. However, one of the variations on the pin field model was to increase the row pitch to 1 mm resulting in longer regions of uncoupled transmission lines; these uncoupled regions are not accounted for by the crosstalk cancellation matrix. It should be noted that power supply paths were modeled as ports so that switching noise is accurately captured when these models are utilized in the complete link simulations. The 3D simulation results were extracted to a frequency-domain, S-parameter model (Touchstone file) solved to 32 GHz.

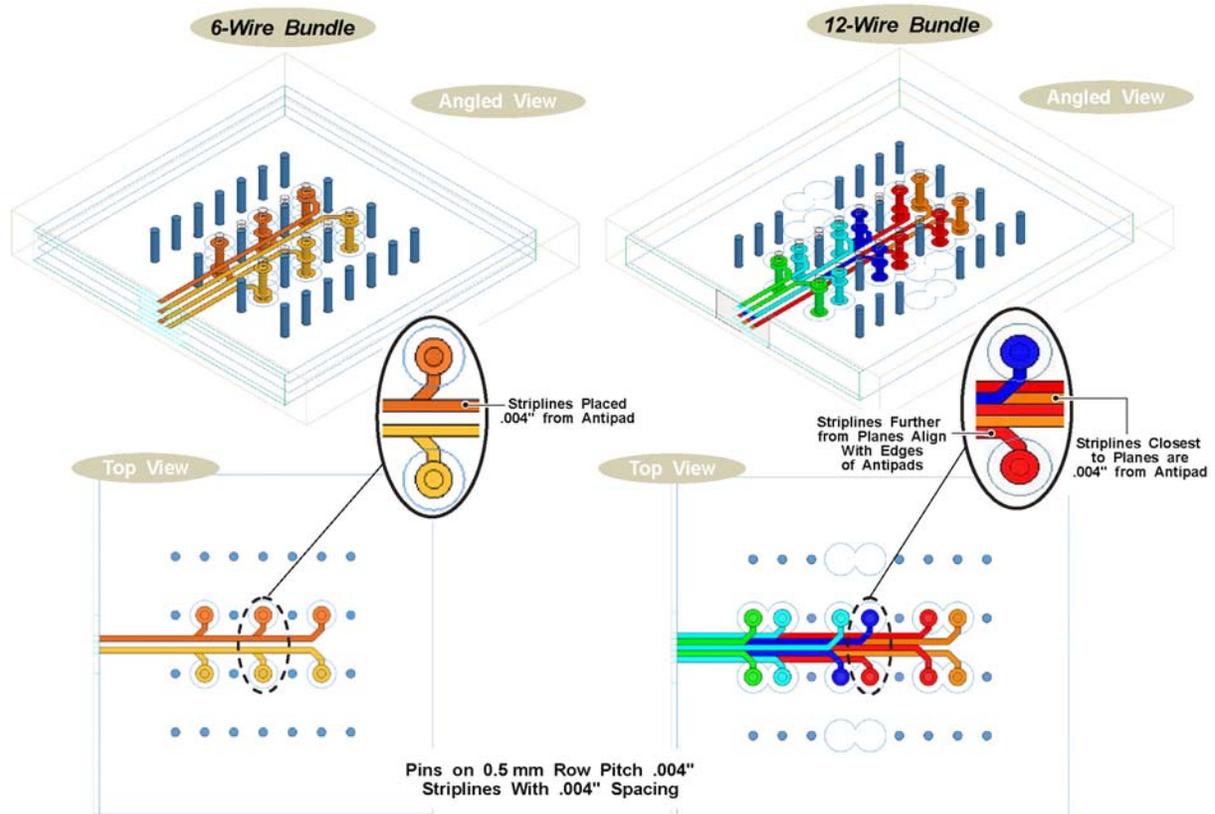

**Figure 5: Via Pin field Models for 6- and 12-Wire Bundles (41504)**

As seen in Figure 5, the aggressive 4 mil wire-wire spacing in the 6-wire bundle is required to maintain a minimum 4 mil clearance from the via antipad in a 1 mm pin pitch design. Typical manufacturing constraints require this clearance to prevent the fabricated transmission lines from overlapping the antipads cut into the reference plane to avoid impedance discontinuities. In the 12-wire bundle design, the same rule is applied to the layers adjacent to the reference planes. However the layers containing three lines are not tightly coupled to the reference plane and can be designed closer (or adjacent to) to the antipad region with minimal impact to the impedance. These two transmission line configurations were designed to provide a realistic comparison for the breakout of a typical 1 mm pin pitch application.

The same pin field models were used at the TX and RX end of the link. Because of the symmetry in the 6-wire bundle, the transmission line lengths could be perfectly matched with careful attention to pin/port assignments in the ADS graphical interface. Unfortunately, the 12-wire bundle could not be perfectly length matched. In the nominal case, the lengths were matched as well as possible. In addition, the trace lengths for both bundle configurations were intentionally mismatched (by varying the nominal pin field implementation) to further examine the effectiveness of crosstalk cancellation with longer uncoupled regions.

Two row pitches (20 mil vs. 40 mil), two length matching options (matched vs. unmatched) and two bundles (6-wire vs. 12-wire) resulted in eight possible pin field configurations. The nominal simulations for both 6- and 12-wire bundles used the 20 mil pin field with the physical lengths matched where possible. Deviations from nominal are noted where applicable.

## Link Architecture: Resistive Termination

The 6-wire bundles were far-end terminated conventionally with 50 Ω resistors to a voltage of $V_{DD}/2$. Using the method described in the theory sections above, a matrix of signal-signal and signal-reference resistors was created to terminate the 12-wire link such that crosstalk in the uniform transmission lines would effectively be cancelled. A baseline matrix was designed with a nearly complete matrix of resistors (per the approach discussed above, though with some of the very high value resistances removed). In addition, two variations of the termination scheme were implemented with a significantly reduced number of resistors in order to quantify the impact of a less constraining design on crosstalk cancellation signaling performance. All resistive matrices are documented in Appendix B.

## Link Architecture: Simulation Stimulus and Patterns

Because of the varying modal impedances for the wires in each bundle, signal-signal logic state relationships will have an impact on signal amplitude at the end of the link. Therefore, three different signal patterns across the 12- and 6-wire buses were explored. Traditional odd- and even-mode signal patterns for the 6-wire bundle, where the signals on a common layer represent the signal pairs, were considered "best" and "worst" case respectively. For the 12-wire bundle, a pseudo-odd- and pseudo-even-mode was assumed as the heuristic "best" and "worst" case pattern, respectively. This description of stimulus patterns is made more clear in Figure 6, where the signal-signal state relationships (including nomenclature) used to stimulate each of the links are illustrated. The random pattern arrangement was meant to emulate a nominal operating condition rather than the pathological "best" and "worst" cases, and thus is expected to perform (with respect to the metrics described below) somewhere in between the two extremes. The crosstalk cancellation SST input buffers were each stimulated with a 16 Gbps, $2^7$-1 PRBS pattern.

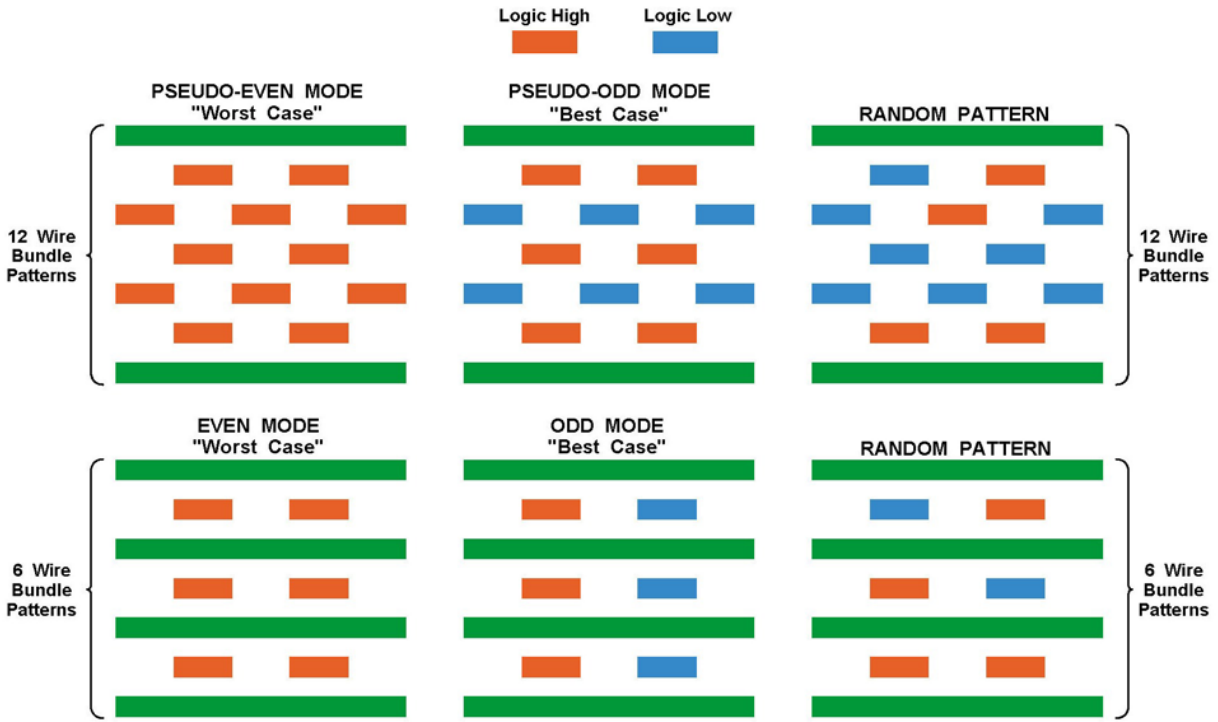

**Figure 6:** Pattern Stimulus Nomenclature for 6- and 12-Wire Bundles; Heuristic Worst and Best Case (41548)

## Simulation Results

As is the case with nearly any signal integrity study of high speed links over complex PWB topologies, a number of factors influence the final performance of a link.  Tight coupling in bus architectures and complex termination networks only exacerbate the analysis challenges.  In this section, results are presented for a number of experiments performed in order to assess crosstalk cancellation signaling performance relative to traditional (single-ended) signaling.

Agilent ADS 2009 for Linux was used for all time-domain simulations.  Custom Matlab scripts were used to capture and measure pertinent eye diagram statistics.  The primary metric by which the two bundle architectures were evaluated was vertical eye diagram opening.  Therefore, other classical eye diagram characteristics, such as jitter, typically evaluated with an eye mask were not thoroughly considered in these simulation results

## Nominal Simulations: Eye Diagrams

Sample eye diagram simulation results are presented in Figure 7.  Wire #3, as documented in Figure 4 for the 12- and 6-wire bundles, was arbitrarily chosen for these example measurements.  The waveforms were evaluated as the difference between the 0.5V reference voltage and the signal amplitude measured at the termination resistor; hence the nominally 0V common-mode voltage observed in these example eye diagrams.

In addition to the custom crosstalk cancelling termination network, the 12-wire bundles were also simulated with conventional to 50 Ω terminations.  This was done with the expectation that the eye would be closed due to the wide-ranging modal impedance in the bundle, but is useful as a simulation benchmark for effectiveness of the crosstalk cancelling termination network.  The expected closed eye diagram was validated by simulation for the "worst" case pattern, which reiterates the need for a more sophisticated termination matrix in a tightly coupled, single-ended application.  Coincidentally, the "best" case pattern used to stimulate the 12-wire bundles resulted in a modal impedance where the nominal, single-ended termination to ground was near 50 Ω.  Hence this is the reason for the optimistic eye opening in the "best" case pattern, 12-wire bundle with 50 Ω terminations.  Further examination of all wires within the bundles under additional pattern stimuli is presented next.

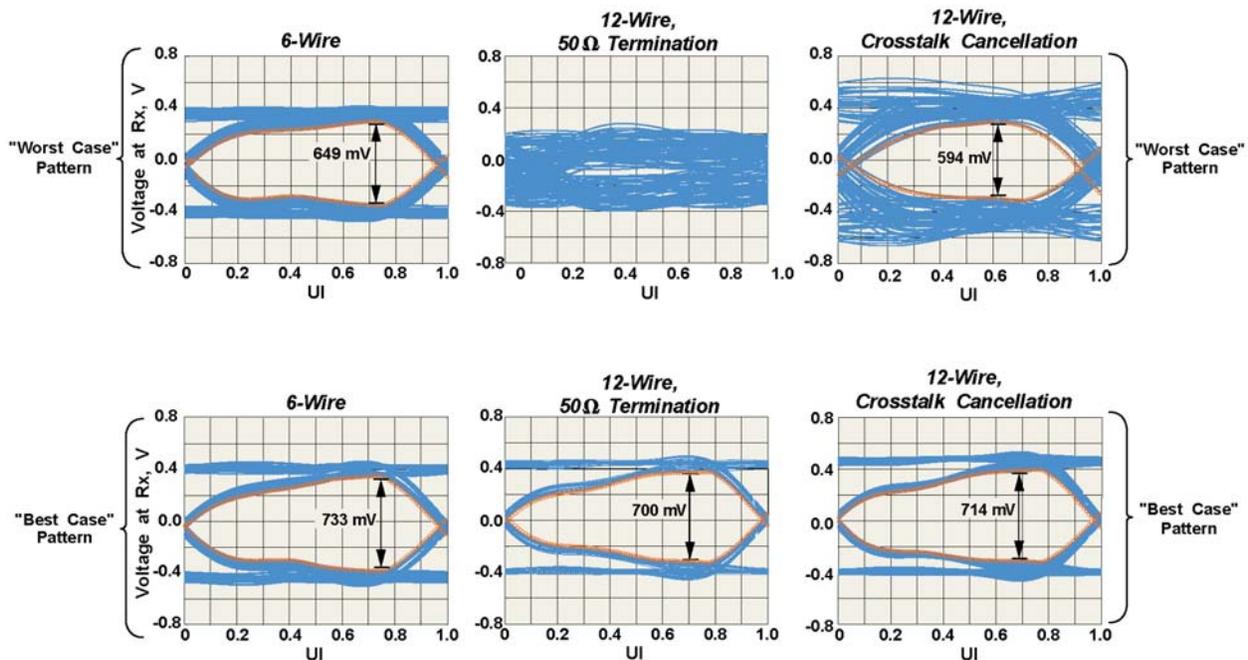

**Figure 7:  Sample Eye Diagrams for 6- and 12-Wire Bundles (Wire #3) Operating at 16 Gbps: 1.67 Ω TX Impedance, Matched Pin field, 20 mil Pin Pitch, "Full" Cancellation Matrix (41547)**

The eye diagram measurements for all wires in each of the 6- and 12-wire simulations are summarized in Figure 8.  In this plot, each small horizontal line represents the vertical eye measurement of a wire within the bundle.  Additionally, markers denoting the minimum, average, and maximum eye opening for each configuration/pattern are shown.  Note that in these plots, it is again apparent that 50 Ω termination is insufficient for the highly coupled 12-wire bundle, given the complete eye closures evident for many of the wires in the bundle, both in the "worst" case and "random pattern" case.

Figure 8 also shows that the "best" and "worst" case pattern experiments bound the performance of the random patterns.  So even though the "worst" case patterns in these experiments were not rigorously determined, these results show that heuristically they are expected to represent "worst" case performance for both the 6-wire and 12-wire bundles regardless of termination specifics.

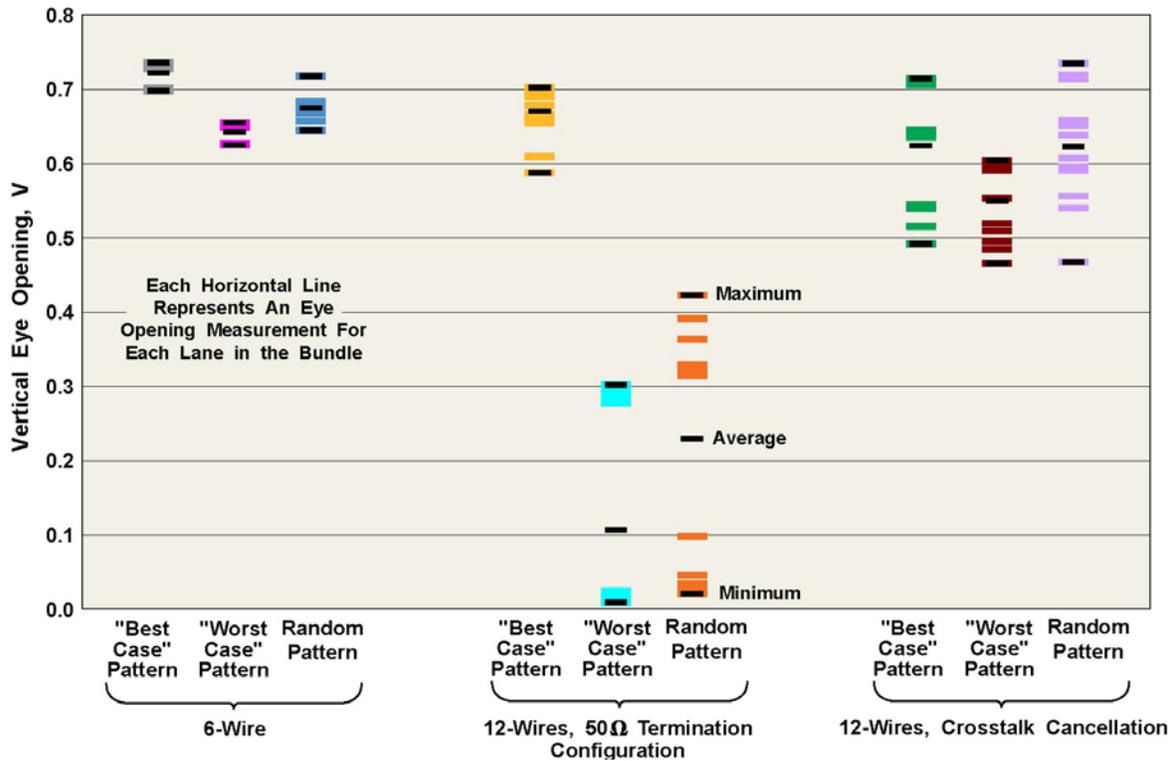

**Figure 8:** Summary of Vertical Eye Diagram Measurements for Each Location in the 6- and 12-Wire Bundles: 16 Gbps, 1.67 Ω TX Impedance, Matched Pin field, 20 mil Pin Pitch, "Full" Cancellation Matrix (41550)

## Simulation Variation: Building the Link and Uncoupled Regions

Even using a simplistic high-speed link model, there were still multiple opportunities for signal degradation in the link. For example, frequency-dependent loss in the transmission lines, non-ideal source impedance, and pin field matching each contribute to eye closure in addition to stimulus pattern effects and the specifics of the termination schemes. By starting with ideal elements and adding each realistic link component one at a time, an understanding of how each element impacts the eye measurements became relatively straightforward. The results of this simulation experiment are provided in Figure 9.

Some interesting observations can be made about the results in Figure 9. The baseline configuration includes no pin field, lossless ideal transmission lines and ideal sources. For the 6-wire bundle, only the "best" case pattern reaches full eye opening. Since the even-mode ("worst" case pattern) impedance is greater than the termination resistance value of 50 Ω, the reflection coefficient is negative thus resulting in eye opening reduction. However, for the case of the 12-wire bundle, the cross-talk termination matrix correctly accounts for the varying modal impedances and hence the eye reaches maximum opening for both "best" and "worst" case stimulus patterns.

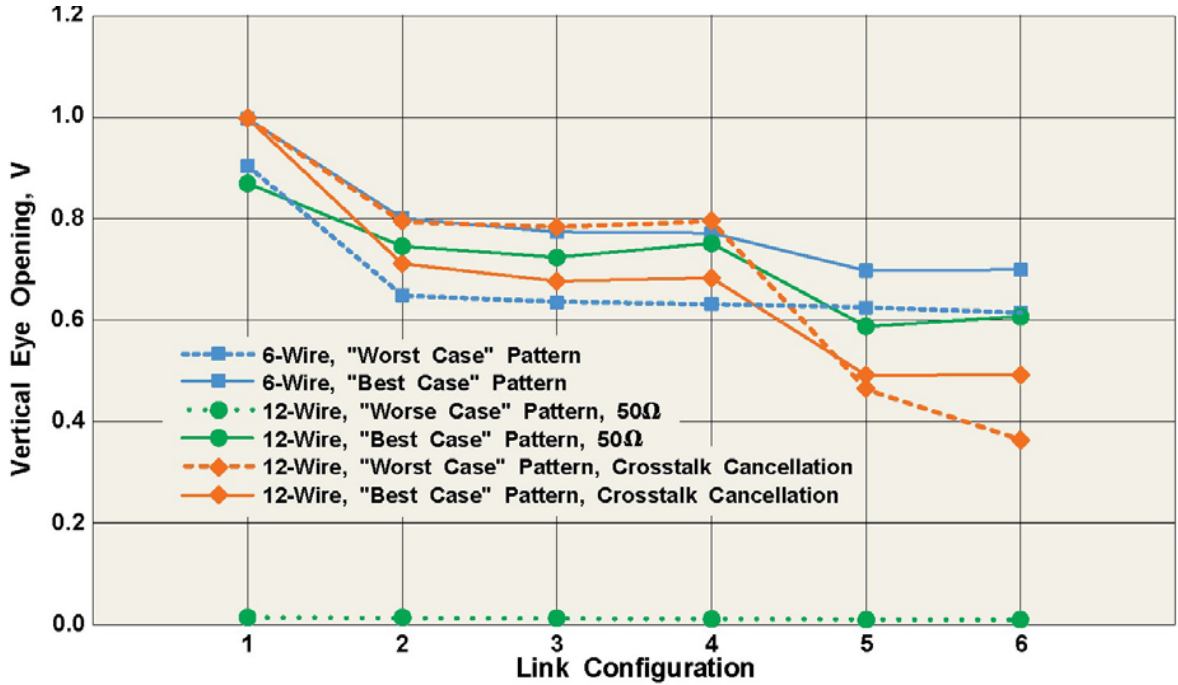

**Figure 9: Link Component Contributions to Vertical Eye Diagram Closure: 20 mil Pin Pitch, "Full" Cancellation Matrix (41549)**

Clearly, the two largest contributors of vertical eye degradation observed in these simulations are the 4" lossy transmission line and the addition of a non-uniform via pin field. In modern high-speed receiver hardware, loss is typically compensated for with equalization. Therefore, while transmission line losses result in significant performance degradation in these simulations, loss is not particularly relevant in the context of this study, as it applies roughly equally to all test cases, and would in practice be readily corrected with equalization.

On the other hand, the large observed performance degradation in Link Configuration 5 in Figure 9 shows that the uncoupled regions in the pin field model are not fully accounted for by the crosstalk cancellation network. This is not unexpected considering the cancellation matrix is designed based on assumptions of a uniform transmission line. The impact of uncoupled length was further examined by increasing the nominal 20 mil pitch pin field model to 40 mils, thereby increasing the amount of uncoupled length. Results comparing the two pin pitches (uncoupled length), both physically matched and unmatched, are shown in Figure 10. The eye continues to close as the uncoupled region is increased.

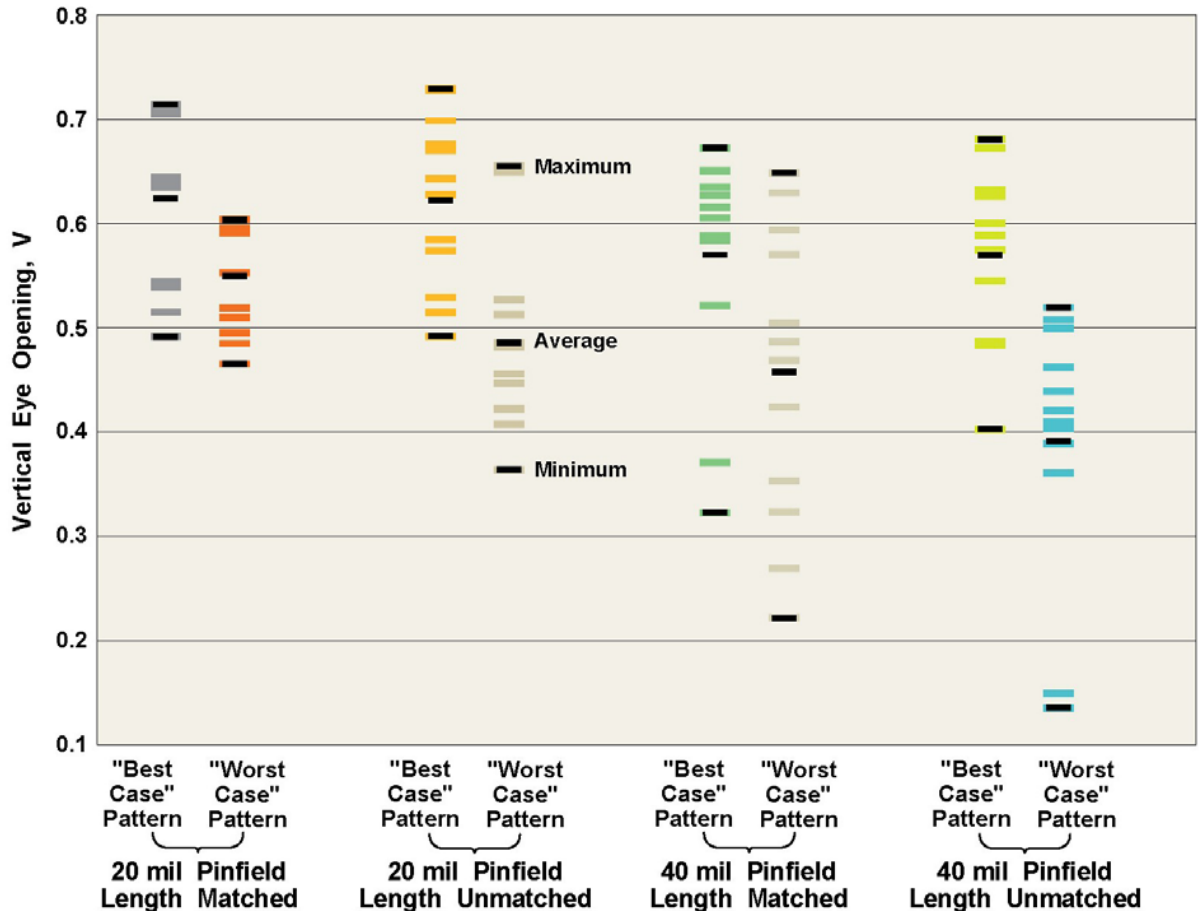

**Figure 10: Impact of Uncoupled Transmission Line (Pin Pitch) on Vertical Eye Opening: 1.67 Ω TX Impedance, "Full" Cancellation Matrix (41554)**

## Simulation Variation: TX Source Impedance

The importance of TX source impedance cannot be understated. Because of the cross-terms in the cancellation matrix, I/O current in each channel of the 12-wire bundle does not simply sink/source to/from the termination voltage. Rather, depending on the logic states across the bus, significant amounts of cross-channel current can terminate into the source impedance of neighboring I/O buffers. The relationship between transmit source resistance and vertical eye opening is illustrated in Figure 11.

There is also a dependence on source impedance for the 6-wire bundle. However, this is a predictable phenomenon due only to the simple voltage division between the source impedance and the single termination resistor. The 12-wire bundle current flow is also predictable based on the logic states of each wire, but as is discussed above, further study would be required to investigate this or other similar possibilities.

As mentioned in the **Link Architecture: Transmit Buffer** section above, Figure 11 drove the selection of 1.67 Ω as the nominal buffer output impedance for the majority of the other simulations. This impedance is related to the physical implementation of the buffer. One copy of the structure results in a nominal 50 Ω impedance, while 50 parallel copies (physically impractical) leads to a 1 Ω impedance. The 12-wire curves in Figure 11 show a saturation in eye opening at low impedances where 1.67 Ω (30 parallel copies of transistors) results in similar performance to 1 Ω but with 40% fewer transistors.

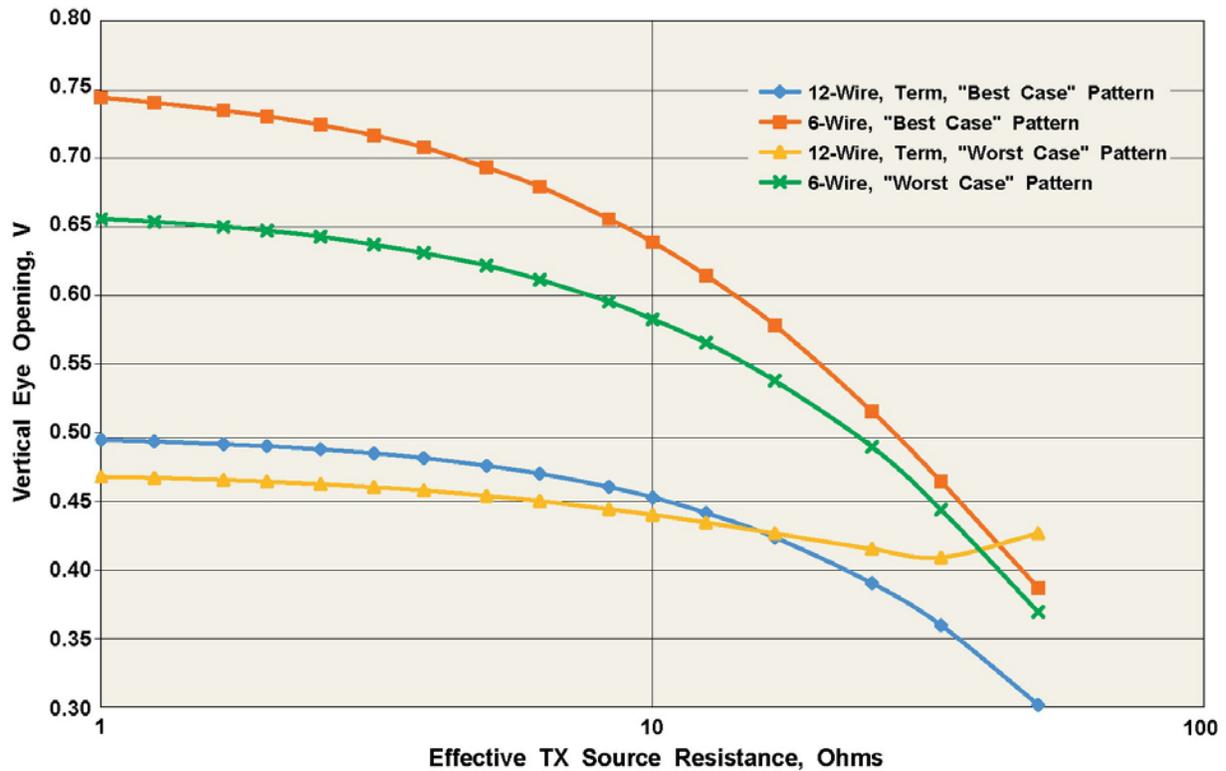

**Figure 11: Impact of Transmitter Source Impedance on Vertical Eye Opening (41551)**

With one exception, the eye opening heights decrease with increasing source resistance. For that exception, which is the case of the 50 Ω transmit impedance for the 12-wire "worst" case pattern in Figure 11, the cause of improvement was not determined. Future work could explore this anomaly more thoroughly, including extending the simulations for all cases beyond 50 Ω to determine if other configurations experience similar behavior. Also, note that in general, the vertical eye opening associated with the 12-wire "worst case" pattern drops off much more slowly with increasing source resistance than does the eye opening for the 6-wire case. In fact, the worst case eye opening for a ~30 Ω buffer in the 12-wire case is roughly equal to that of the original 50 Ω buffer in the 6-wire case.

When further consideration is given to designing an optimal buffer for crosstalk cancellation signaling, the horizontal eye opening must also be examined since timing closure (jitter) due to inter-symbol interference (ISI) has not been carefully considered in this study.

## Simulation Variation: Crosstalk Cancellation Matrix

The number of resistors required in the cancellation matrix was also studied as it impacts the practicality of physical implementation. Obviously, minimizing the integrated circuit real estate required to implement the termination network would be beneficial to the realization of the crosstalk cancellation signaling concept. The original network, termed "Full", was created using an arbitrary maximum limit of 60 kΩ for self terms and 120 kΩ for cross terms in the termination matrix, resulting in 66 total resistors. See Appendix B for further discussion and documentation of resistor values. A secondary matrix was created, referred to as "Reduced", by changing the arbitrary maximum resistor value limits to 500 Ω and 1000 Ω for self and cross terms respectively. This results in 50% fewer resistors; 33 total. Finally, a third reduction was performed with an arbitrary 300 Ω and 600 Ω self and cross term cutoff resulting in 26 total resistors; a 67% reduction. In this implementation, only the nearest neighbor resistors remain. Further reduction would create floating signal lines and obviously poor performance results.

The link simulations were performed using both of the "Reduced" matrices and compared to the "Full" network in Figure 12. Generally, there was little observable performance impact when the reduced matrices were used. This is very encouraging for the prospect of future implementation.

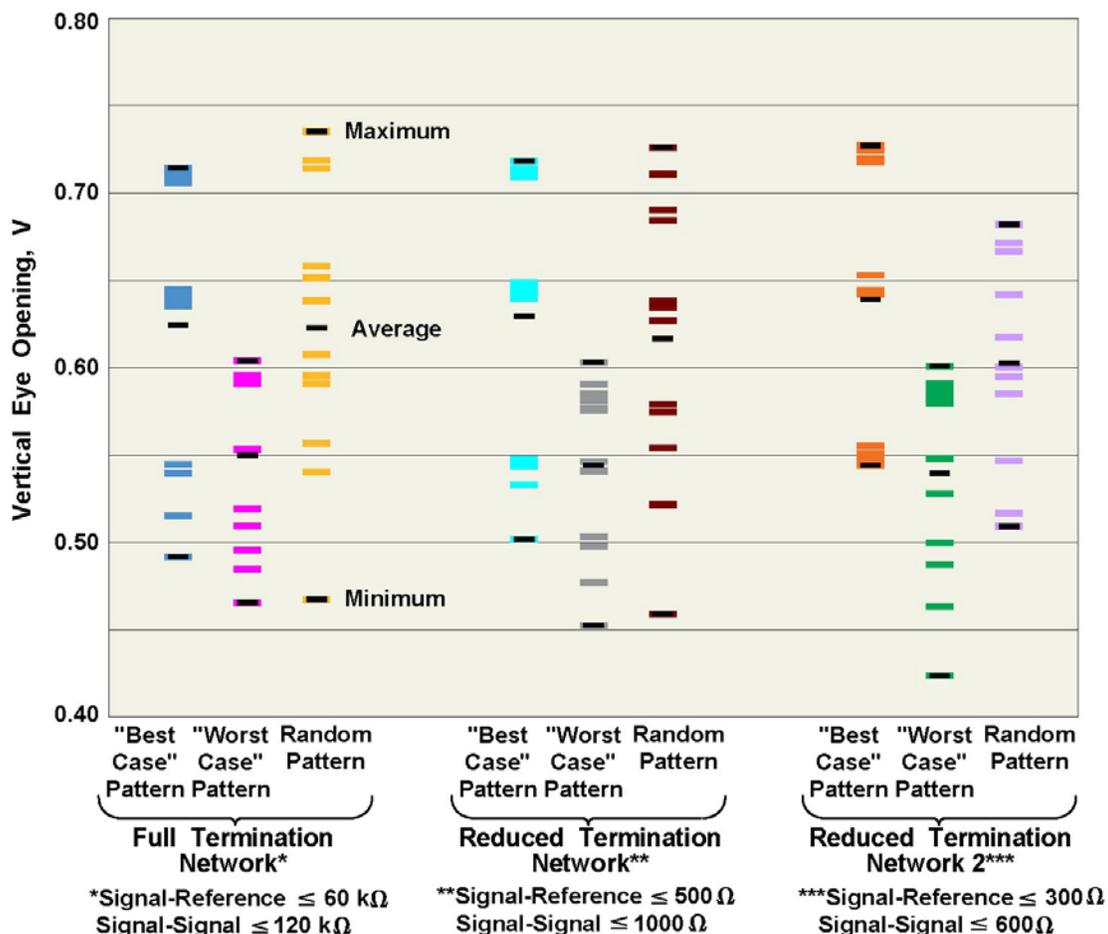

**Figure 12: Impact of Reduced Crosstalk Cancellation Network Resistors on Vertical Eye Opening (41552v2)**

**Crosstalk Cancellation Signaling Conclusions**

The simulation results presented in this paper suggest that using crosstalk cancellation signaling in a tightly coupled, single-ended application can produce signal quality approaching conventional signaling but with roughly twice the signal density. There are a few caveats to this conclusion that should be explored further.

For example, it was presented that uncoupled channel length, unaccounted for by the crosstalk cancellation matrix in these experiments, does reduce the signal quality relative to the length of the uncoupled region. Potentially, the method presented above could be extended to non-uniform media which include uncoupled regions, allowing signal quality to be optimized with variations on the simplistic termination matrix values. This extension of the work must be especially considered for more complex links such as those regularly employed in backplane applications containing many other non-uniform components and discontinuities. It is hoped that these methods allow signal quality improvements to continue to be observed, even for more complex link architectures.

Additional examination of the PWB cross-sections in Figure 3 indicates the potential for other implementation-related non-idealities to the uniform channel. For example, manufacturing tolerances on PWB layer-layer registration could result in shifting of the relative trace locations within a bundle. This non-ideality is not accounted for by a static cancellation matrix that was designed specifically for the as-designed RLGC matrix of the wire bundle in the nominal case. Issues such as this will require an adjustable resistance matrix in the receiver that can respond to a training sequence where the values are optimized for the channel as it has been manufactured; rather than how it was designed.

Some time was required to become familiar with crosstalk cancellation signaling principles of operation and to develop custom diagnostic tools such as the Matlab code to predict switching and power figures of merit. With this better understanding, it may be possible to devise improved wire bundle designs, pin field breakout definitions and related design aspects to further exploit the benefits of crosstalk cancellation signaling.

Perhaps the single most important issue to address is the suggested transmit buffer. This simulation work made use of the SST voltage source driver, but as was noted, the spread in effective modal impedances results in a pattern-dependent voltage division, bringing about eye height reduction. Although the simulated eye openings in many of the cases considered here might still be sufficient with a modest sized SST buffer, it seems investigation of alternate buffer schemes, such as the use of current mode drivers would be one of the most important next steps.

On a related note, both buffers and receivers used in crosstalk cancellation signaling should be studied to determine the effectiveness of incorporating equalization methodologies. In addition to the wide ranging modal impedances, the wire bundles also exhibit varying modal propagation constants.

# Appendix A: Crosstalk Cancellation Matrix Termination Calculations

Numerous papers discuss computation of the characteristic impedance matrix from a uniform coupled transmission line. Using methods borrowed from Lei [7], Schutt-Aine [1] and others the equations below were utilized for our work.

Solve for eigen values and eigen vector of the inverse of the L, the inductance matrix

$$L^{-1} \bullet l_{vec} = l_{val} \bullet l_{vec}$$

Compute the S matrix

$$S = l_{vec} l_{lval}^{-1/2}$$

Define M matrix

$$M = S' \bullet C \bullet S$$

Solve for eigen values and eigen vector of the M

$$M \bullet m_{vec} = m_{val} m_{vec}$$

Define Mv matrix

$$M_v = S \bullet m_{vec} \bullet m_{val}^{-1/4}$$

Define Mi matrix

$$M_i = M_v^{-1}$$

Define Zc matrix

$$Z_c = M_v \bullet M_v$$

# Appendix B: Detailed Crosstalk Cancellation Simulation Parameters

Three crosstalk cancellation resistance matrices, referred to as "Full", "Reduced" and "Reduced 2", were evaluated during this study. Each of these matrices are documented in Table 1. In each model, a limit was applied to the resistor values included in the matrix. For the "Full" model, the limit for signal-signal and signal-reference was 120 kΩ and 60 kΩ respectively. For the "Reduced" model, the limit for signal-signal and signal-reference was 1000 Ω and 500 Ω respectively. For the "Reduced 2" model, the limit for signal-signal and signal-reference was 600 Ω and 300 Ω respectively. All resistors documented in Table 1 represent the "Full" model (66 resistors). The values highlighted in **bold** and *italics* were used in the "Reduced" matrix (33 resistors). The values highlighted only in *italics* were used in the "Reduced 2" matrix (26 resistors). All values are defined in Ohms.

| | | | | | |
|---|---|---|---|---|---|
| *R1_1* | *110.85721* | | | | |
| **R1_2** | **663.03745** | R2_2 | 110.85727 | | |
| *R1_3* | *252.17476* | R2_3 | 17371.80328 | *R3_3* | *224.54748* |
| *R1_4* | *284.86464* | R2_4 | 284.86474 | **R3_4** | **642.14998** |
| R1_5 | 17371.75169 | R2_5 | 252.17483 | R3_5 | NA |
| R1_6 | 2017.01452 | R2_6 | 20628.38651 | *R3_6* | *249.31413* |
| R1_7 | 20628.10818 | R2_7 | 2017.00556 | R3_7 | 17101.09832 |
| R1_8 | 15324.10950 | R2_8 | NA | *R3_8* | *568.14039* |
| R1_9 | 79597.11234 | R2_9 | 79606.24676 | R3_9 | 14788.81149 |
| R1_10 | NA | R2_10 | 15323.60895 | R3_10 | NA |
| R1_11 | NA | R2_11 | NA | R3_11 | 15318.46175 |
| R1_12 | NA | R2_12 | NA | R3_12 | NA |
| | | | | | |
| R4_4 | 1417.93431 | | | | |
| **R4_5** | **642.15035** | *R5_5* | *224.54736* | | |
| *R4_6* | *284.48164* | R5_6 | 17101.09114 | R6_6 | 1941.99281 |
| *R4_7* | *284.48160* | R5_7 | 249.31422 | **R6_7** | **656.24375** |
| R4_8 | 14789.49541 | R5_8 | NA | R6_8 | 249.30971 |
| R4_9 | 2108.17695 | R5_9 | 14788.58366 | R6_9 | 284.48129 |
| R4_10 | 14789.61216 | *R5_10* | *568.14056* | R6_10 | 17101.16810 |
| R4_11 | 79603.82438 | R5_11 | NA | R6_11 | 2016.97037 |
| R4_12 | 79601.53980 | R5_12 | 15318.32199 | R6_12 | 20629.41956 |
| | | | | | |
| R7_7 | 1942.00554 | | | | |
| R7_8 | 17101.36652 | *R8_8* | *224.59869* | | |
| *R7_9* | *284.48133* | **R8_9** | **642.13988** | R9_9 | 1417.92733 |
| *R7_10* | *249.30967* | R8_10 | NA | **R9_10** | **642.14046** |
| R7_11 | 20628.62871 | *R8_11* | *252.17010* | *R9_11* | *284.86731* |
| R7_12 | 2016.97112 | R8_12 | 17372.40969 | *R9_12* | *284.86730* |
| | | | | | |
| *R10_10* | *224.59879* | | | | |
| R10_11 | 17372.18870 | *R11_11* | *110.84691* | | |
| *R10_12* | *252.17002* | **R11_12** | **663.07539** | *R12_12* | *110.84690* |

**Table 1: 12-Wire Resistive Termination Resistance Values**

25